\documentclass[10pt, conference, compsocconf]{IEEEtran}

\IEEEoverridecommandlockouts

\usepackage{cite}
\usepackage{amsmath,amssymb,amsfonts}
\usepackage{algorithm}
\usepackage{algpseudocode}
\usepackage{graphicx}
\usepackage{textcomp}
\usepackage{xcolor}
\usepackage{caption}
\usepackage{subcaption}
\usepackage{multirow}
\usepackage{balance}
\usepackage{nopageno}
\usepackage{amssymb} 
\usepackage{adjustbox} 
\usepackage{array}
\usepackage[left=1.62cm,right=1.62cm,top=0.72in]{geometry}

\def\BibTeX{{\rm B\kern-.05em{\sc i\kern-.025em b}\kern-.08em
    T\kern-.1667em\lower.7ex\hbox{E}\kern-.125emX}}
\begin{document}

\title{LLMAC: A Global and Explainable Access Control Framework with Large Language Model
~\vspace{-15pt}
}

\author{
\IEEEauthorblockN{\textsuperscript{} Sharif Noor Zisad}
\IEEEauthorblockA{\textit{Department of Computer Science} \\
\textit{University of Alabama at Birmingham}\\
Birmingham, Alabama, USA \\
szisad@uab.edu
~\vspace{-40pt}
}
\and
\IEEEauthorblockN{\textsuperscript{}Ragib Hasan}
\IEEEauthorblockA{\textit{Department of Computer Science} \\
\textit{University of Alabama at Birmingham}\\
Birmingham, Alabama, USA \\
ragib@uab.edu
~\vspace{-40pt}
}
}

\maketitle

\thispagestyle{plain}
\pagestyle{plain}

\begin{abstract}
\label{sec:abstract}
Today's business organizations need access control systems that can handle complex, changing security requirements that go beyond what traditional methods can manage. Current approaches, such as Role-Based Access Control (RBAC), Attribute-Based Access Control (ABAC), and Discretionary Access Control (DAC), were designed for specific purposes. They cannot effectively manage the dynamic, situation-dependent workflows that modern systems require. In this research, we introduce LLMAC, a new unified approach using Large Language Models (LLMs) to combine these different access control methods into one comprehensive, understandable system. We used an extensive synthetic dataset that represents complex real-world scenarios, including policies for ownership verification, version management, workflow processes, and dynamic role separation. Using Mistral 7B, our trained LLM model achieved outstanding results with 98.5\% accuracy, significantly outperforming traditional methods (RBAC: 14.5\%, ABAC: 58.5\%, DAC: 27.5\%) while providing clear, human readable explanations for each decision. Performance testing shows that the system can be practically deployed with reasonable response times and computing resources. 
\end{abstract}
\vspace{-5pt}
\section{Introduction}
\label{sec:introduction}
Access control is a key part of information security. It provides mechanisms that determine who can access specific resources and perform certain actions within a system. At its core, it protects information and assets by ensuring only authorized individuals can interact with protected resources according to set policies and procedures~\cite{benantar2006access}. Over the years, access control has evolved from simple password-based systems to advanced frameworks that address the complex security needs of modern, distributed, and dynamic enterprises~\cite{qiu2020survey}.

Traditional access control models have been developed to meet organizational needs and security goals. Role-Based Access Control (RBAC) was developed to reduce administrative overhead by organizing permissions around job functions and roles, making it effective in large enterprises with clear hierarchies~\cite{sandhu1998role}. It simplifies management by grouping users with similar responsibilities and assigning permissions collectively. Attribute-Based Access Control (ABAC) allows more detailed policy expression by evaluating multiple attributes of users, resources, and environments. This enables fine-grained decisions based on context, such as time, location, or data sensitivity~\cite{hu2015attribute}. Discretionary Access Control (DAC) provides resource owners the authority to set access permissions, providing flexibility in collaborative environments with changing data sharing needs~\cite{li2005safety}. While each model meets specific needs well, modern enterprises increasingly require solutions that combine the strengths of multiple approaches and adapt to rapidly changing business demands~\cite{aftab2020hybrid}.

Contemporary organizations face major challenges that show the limitations in traditional access control frameworks~\cite{kayes2020survey}. These systems struggle with dynamic workflows where access decisions must consider resource states, time constraints, and multi-step approvals that change with business context and regulations~\cite{moharir2025contextual}. Modern enterprises need policies that handle complex dependencies, evaluating multiple interconnected factors at once, such as user credentials, device security, location, and real time risk~\cite{ohanekwu2025context}. The administrative burden of maintaining these systems grows rapidly as organizations scale, adding significant overhead in creating, debugging, and updating access rules across diverse environments~\cite{KOKILA2025100057}.

Large Language Models (LLMs) can help address access control challenges since they are good at recognizing patterns, understanding context, and using natural language~\cite{wu2025llm}. Unlike traditional systems that require rules for every situation, LLMs can learn from examples and apply that knowledge to new cases while following company policies. Their ability to process and combine information from multiple sources allows them to handle the complex decision making which is needed in modern access control~\cite{yamin2024applications}. Additionally, LLMs can provide human understandable explanations for their decisions, meeting the explainability requirements that traditional systems often lack~\cite{zisad2023evolutionary}. This makes access control more transparent, enabling the system to communicate its reasoning to users, administrators, and auditors in natural language~\cite{bhattacharjee2023towards}.

In this research, we present a new LLM-based framework that unifies RBAC, ABAC, and DAC into one unified system, keeping the strengths of each model while addressing their weaknesses. Using a synthetic dataset that reflects real world organizational workflows, we show that our approach achieves higher accuracy and better usability than traditional methods. Unlike rule-based systems, our framework learns from policy examples, reducing errors and closing security gaps. It also provides clear explanations for its decisions, improving transparency, accountability, and compliance. The adaptive nature of LLMs allows the system to evolve with new examples and threats, helping organizations enhance security over time. By consolidating multiple access control models into one framework, our approach reduces complexity, lowers administrative effort, and creates a consistent, more secure infrastructure. The LLM learns from structured access control examples to identify patterns in access decisions, applies this knowledge to new situations, and clearly explains its decision.

\noindent\textbf{Contribution:}
The contributions of this paper are as follows:
\begin{enumerate}
    \item We proposed an LLM-based unified access control framework combining traditional models.
    \item We demonstrated its accuracy, adaptability, and explainability compared to traditional models through a use case.
\end{enumerate}

\noindent\textbf{Organization:}
The structure of this paper is as follows: Section \ref{sec:introduction} covers the introduction of interaction provenance and access control models. Section \ref{sec:background} discusses the background, working principle, and limitations of these models. Related works are reviewed in Section \ref{sec:literature_review}, while we introduce the architecture of LLMAC in Section \ref{methodology}. Section \ref{sec:results} presents the experimental results. Finally, Section \ref{sec:conclusion} concludes the paper with some future directions.\\
\vspace{-5pt}
\vspace{-15pt}
\section{Background}
\label{sec:background}
\subsection{Traditional Access Control Models} 
An Access Control Model is a framework that defines how permissions are granted to users and systems to access resources within a computing environment \cite{sandhu1994access}. There are various models, including Role-Based Access Control (RBAC) \cite{sandhu1998role}, Attribute-Based Access Control (ABAC)~\cite{hu2015attribute}, Discretionary Access Control (DAC)~\cite{li2005safety}, and History-Based Access Control (HBAC)\cite{edjlali1998history}, each designed to enforce policies on who can access what based on predefined rules.

\subsubsection{Role-Based Access Control (RBAC)} is one of the most widely used access control models in modern computing environments~\cite{sandhu1998role}. In RBAC, access to resources is determined by the roles assigned to users, rather than their individual identities. Each role is associated with a set of permissions, and users are granted access to resources based on the roles they assume within an organization. For instance, a user assigned the role of "Manager" might have the ability to approve documents and view sensitive data, while a "Staff" role might be restricted to submitting reports without viewing confidential information. 

Although Role-Based Access Control (RBAC) offers simplicity and scalability, it has several limitations. One major issue is the explosion of roles ~\cite{ahn2000role}, where the number of roles becomes unmanageable in large organizations as more specific roles are created to accommodate diverse access needs. RBAC also lacks flexibility ~\cite{li2006security}, as it cannot easily handle dynamic or context-sensitive access requirements, such as time-based access or access based on the current location of a user. Additionally, the model assumes static roles and permissions, which may not reflect the evolving nature of users' responsibilities or tasks. This rigidity can lead to overprovisioning or underprovisioning of permissions, increasing the risk of unauthorized access or hindering productivity. 
\subsubsection{Attribute-Based Access Control (ABAC)} is a flexible and dynamic access control model that grants or denies access based on attributes associated with users, resources, actions, and the environment~\cite{hu2015attribute}. In ABAC, decisions are made by evaluating these attributes against defined policies, which can include factors such as user roles, resource types, time of access, location, or device being used. Unlike RBAC, ABAC allows for more granular and context-aware access control by considering a wide range of attributes. This flexibility makes ABAC well suited for environments where access decisions depend on multiple variables or need to change dynamically based on the situation. For instance, in a hospital setting, a doctor may access patient records only during working hours and within the hospital premises. 

Although ABAC provides a higher level of control, its complexity can be a challenge ~\cite{servos2017current}. Defining and managing the necessary attributes and policies can be difficult, especially in large-scale systems with diverse access needs. However, ABAC's ability to handle complex, real-time scenarios makes it a powerful tool in environments including cloud computing, healthcare, and government, where access must be continuously adapted to changing conditions and requirements.

\subsubsection{Discretionary Access Control (DAC)} is an access control model in which the owner of a resource, typically a file or system, has the discretion to decide who can access it and what actions they can perform~\cite{li2005safety}. This model allows users to grant or revoke permissions for other users or groups based on their own judgment. DAC permissions are usually based on access control lists (ACLs)~\cite{qian2001acla}, where the owner specifies which users or processes can read, write, or execute the resource.

\begin{table*}[!ht]
\caption{Traditional Access Control Models}
  \label{tab:traditional_models}
\centering
\begin{tabular}{|p{4cm}|p{5cm}|p{3.5cm}|p{4cm}|}
\hline
\textbf{Access Control Model} & \textbf{Working Principle} & 
\textbf{Pros} & \textbf{Cons} \\
\hline
Role-Based Access Control (RBAC) & Access is determined by the roles assigned to users, with each role associated with a set of permissions. Users are granted access based on their roles within an organization. & Simplicity, scalability, and ease of administration in large organizations. & Role explosion, lack of flexibility for dynamic or context-sensitive access, and complex role management over time. \\
\hline
Attribute-Based Access Control (ABAC) & Access is granted based on attributes related to users, resources, actions, and environmental conditions, evaluated against defined policies. & Provides fine-grained, context-aware access control, suitable for dynamic scenarios and large-scale systems. & Complexity in managing attributes and policies, especially in diverse, large systems. \\
\hline
Discretionary Access Control (DAC) & The resource owner decides who can access the resource and what actions they can perform, typically via Access Control Lists (ACLs). & Flexibility for resource owners to manage access permissions. & Inconsistent security, risk of unauthorized access, lack of centralized management, hard to enforce uniform policies. \\
\hline
\end{tabular}
\vspace{-10pt}
\end{table*}

The primary problem with Discretionary Access Control (DAC) lies in its dependence on resource owners to manage permissions ~\cite{downs1985issues}, which can lead to inconsistent or insecure access policies. Since individual users have full control over their resources, they may unintentionally grant access to unauthorized users or fail to apply sufficient restrictions, creating security risks. This flexibility, while convenient, can result in data leaks or malicious exploitation if permissions are not carefully monitored. In addition, DAC lacks centralized oversight ~\cite{thomas1995discretionary}, making it difficult for organizations to enforce uniform access control policies or to audit and track access throughout the system. The difference between these traditional access control models is described in Table \ref{tab:traditional_models}.

\subsection{LLM for Access Control}
Large Language Models (LLMs) are neural networks trained on large datasets to predict token sequences, enabling them to understand instructions, reason over inputs, and generate context aware outputs~\cite{alberts2023large}. They encode inputs, such as users, actions, resources, and context, into high-dimensional representations~\cite{bopp2024case}. Using transformer based attention, they capture relationships and produce outputs that can be formatted for specific purposes, such as a JSON decision with an explanation. When finetuned on access decisions and rationales, LLMs can learn policies that address complex conditions, including workflow states, temporal rules, and cross-entity relationships~\cite{lin2024data}. Techniques like parameter-efficient finetuning~\cite{fu2023effectiveness} and quantization~\cite{bai2023unified} let them adapt to organizational needs with limited resources, while structured output or deterministic decoding ensures consistent, auditable results. By combining multiple signals and offering clear explanations, LLMs function both as enforcement engines and explanation generators, aligning access decisions with human-understandable reasoning.

In access control, LLMs can be a unified decision layer that can combine the benefits of RBAC, ABAC, and DAC without requiring administrators to choose a single approach. Policies may be encoded as structured prompts or learned from labeled examples. It will allow the model to handle complex, context-dependent conditions while providing clear explanations~\cite{ADANZA2025111647}. Security and privacy can be ensured through measures such as filtering sensitive inputs, enforcing least privilege on contextual data, constraining outputs to comply with policies, and validating decisions before enforcement. Additional protections, including model isolation per tenant, encrypted logs, controlled access to prompts and traces, rate limiting, and continuous testing, can reduce risks~\cite{10.1145/3712001}. These combined strategies allow LLMs to manage dynamic and intricate access policies effectively, while maintaining transparency and robust security for modern enterprise systems.
\section{Literature Review}
\label{sec:literature_review}
In this section, we will discuss the recent works that combine access control with modern AI. The focus will be on large language models, policy design, and clear explanations of how decisions are enforced.

Rubio et. al. developed a model that enforces access control policies with LLMs \cite{rubio2024pairing}. They proposed a pipeline that turns high-level policy goals into clear decisions which helps to reduce mistakes when interpreting policies. It uses formal policy descriptions in the LLM’s context so the decisions can be traced back to the rules, making them easier to explain and check in sensitive systems. However, it depends on the quality of prompts and the accuracy of formal rules, which may cause problems if the rules are missing or outdated.

Lawal et al. (NIST/UTSA) introduced a method to change natural language policy text into structured access rules using in-context prompting instead of complex fine-tuning \cite{lawal2024translating}. Their approach extracts key parts such as subjects, actions, attributes, and constraints from documents like compliance rules, user stories, and SOPs, then converts them into a graph and policies for consistent enforcement. The drawback is the results depend on prompt quality and document differences, which may lower consistency unless extra checking and corrections are added.

Vatsa et al. explored how LLMs can create access control policies from plain language descriptions of what is allowed and denied, without needing fine tuning~\cite{vatsa2025synthesizing}. Their study shows that using structured, syntax aware prompts helps the model generate more accurate and checkable policies than simple prompts. The main limitations is the results depend heavily on prompt design, and need stronger support across different policy languages and large real-world systems.

Mai et al. proposed a method for fine-grained ABAC policies by using multiple LLMs, retrieval-augmented prompts, and priority optimization \cite{mai2025llm}. Their workflow provides detailed ABAC rules, improves them with extra context. It has high flexibility, better handling of unusual cases, and a clear process for moving generated policies into real decision systems. However, there is a chance of conflicts when many rules are created. It also dependence on accurate ground-truth data for priority optimization over time.

Based on the discussion on previous works, we can see they only shape policies with LLMs. There is a lack of a full system with stable explanations. Our model, LLMAC, provides a unified and explainable framework that combines RBAC, ABAC, and DAC, with clear evaluations and practical least privilege use.

\vspace{-5pt}
\section{Methodology}
\label{methodology}
\subsection{System Architecture}
\label{sec:system_architecture}
\begin{figure}[!ht]
    \includegraphics[width=1\columnwidth]{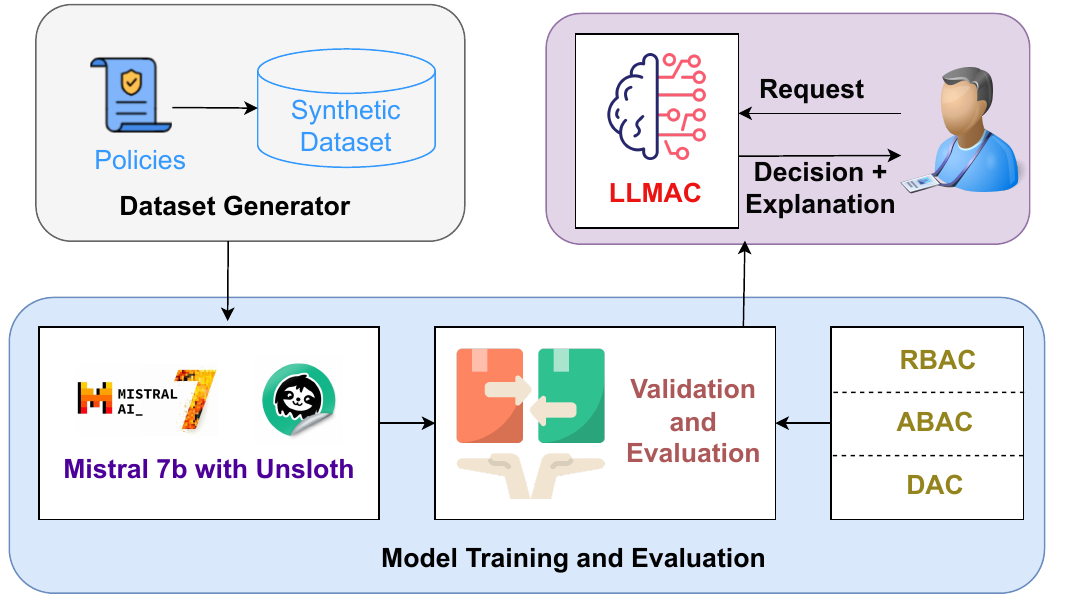}
    \centering
    \caption{System Architecture}
    \label{fig_system_architecture}
    \vspace{-15pt}
\end{figure}

The architecture of our proposed model, LLMAC (LLM-based Access Control) consists of three interconnected modules that work together to provide a unified, explainable access control framework. The architecture of the system is presented in Figure \ref{fig_system_architecture}.

The foundation of the system begins with the Dataset Generator, which transforms organizational Policies into a Synthetic Dataset. This module takes formal policy specifications (including RBAC roles, ABAC attributes, DAC ownership rules, and complex workflow constraints) and systematically generates diverse access control scenarios. Each generated example includes user identities, requested actions, target resources, contextual attributes, ground-truth access decisions (allow/deny), and human-readable explanations that trace back to the governing policy clauses. This synthetic approach ensures comprehensive coverage of edge cases and policy combinations that might be rare in real-world logs.

The synthetic dataset flows into the model training and evaluation module, which performs two important functions. First, it fine-tunes a Mistral 7B language model using the Unsloth framework~\cite{zheng2024llamafactoryunifiedefficientfinetuning}, applying parameter-efficient techniques like LoRA (Low-Rank Adaptation) to adapt the pre-trained model to the organization's specific access control policies~\cite{10901572}. The model learns to map structured authorization requests to binary decisions while generating explanatory rationales. Second, this module implements baseline access control methods (RBAC, ABAC, DAC) as comparison systems and conducts comprehensive validation and evaluation across all approaches. This includes measuring accuracy, precision, recall, F1 scores, operational metrics like latency and throughput, and qualitative assessment of explanation quality and policy alignment

The trained model is then deployed in the core LLMAC runtime system, which handles live authorization requests. A User (or application acting on behalf of a user) submits a request containing the essential authorization parameters, such as user identity, requested action, target resource, and relevant contextual information. The system processes this request through the trained language model, which evaluates the input against learned policy patterns to generate a decision (allow or deny) with a clear explanation that describes the reasoning behind the decision. This explanation references specific policy conditions, workflow states, or constraints that influenced the outcome, providing transparency for auditing and helping users understand why access was granted or denied.

\subsubsection{Experimental Dataset}
In this experiment, we used a comprehensive set of access control policies originally defined in Park et al.~\cite{park2012provenance} to govern a class management system. These policies are carefully designed to enforce fine-grained and workflow-aware authorization decisions, ensuring strict content protection and privacy protections for homework submissions and reviews. Using multiple access control paradigms including origin based controls, versioning mechanisms, dynamic workflow state dependencies, and separation of duty principles, these policies enable the system to handle contemporary security challenges prevalent in collaborative, distributed academic environments. Table \ref{tab:policies} summarizes the access control policies.

\begin{table}[h!]
  \centering
  \caption{Summary of Access Control Policies}
  \label{tab:policies}
  \begin{tabular}{|p{0.8cm}|p{1.5cm}|p{2.5cm}|p{2.5cm}|}
    \hline
    \textbf{Policy ID} & \textbf{Action} & \textbf{Key Conditions} & \textbf{Security/Privacy Impact} \\
    \hline
    P1 & Upload Homework & Any legitimate user (student) can upload assignments. No user or state constraints. & Open access, Low risk, Ensures usability. \\
    \hline
    P2 & Replace Homework & Authors may replace an unsubmitted homework version. & Prevents version abuse, Enforces origin and version integrity. \\
    \hline
    P3 & Submit Homework & Only the original author may submit if not already submitted. & Origin based authentication, Mitigates impersonation threat. \\
    \hline
    P4 & Review Homework & Must be a submitted homework, Reviewer is neither author nor previous reviewer, Review count$<$3, Assignment ungraded. & Dynamic separation of duty, Mitigates collusion, Ensures workflow integrity and privacy of review process. \\
    \hline
    P5 & Revise Review & Only the review creator may revise before grading. & Preserves data provenance and privacy, Restricts unauthorized tampering. \\
    \hline
    P6 & Grade Homework & Grading only after at least 2 reviews exist. & Reduces risk of biased or premature grading, Enforces workflow compliance. \\
    \hline
    P7 & Append Review to Grade & Only the grade creator can append matching review. & Ensures linkage between grade and review with fine grained auditability, Enhances traceability for privacy and compliance. \\
    \hline
  \end{tabular}
  \vspace{-5pt}
\end{table}

The policies effectively model the complexity of real world scenarios where multiple stakeholders such as students, reviewers, and graders, interact with shared digital resources (homework assignments, peer reviews, grades) across distinct lifecycle stages, such as creation, submission, iterative review, and final grading. By embedding specific, dynamically evaluated conditions based on system state, user roles, and relationships, and workflow progression. These policies provide robust security against emerging threats, including unauthorized data modification, impersonation, and privilege escalation. The approach emphasizes transparency and explainability by encoding decision logic that facilitates traceability, aligning with modern regulatory and compliance requirements for secure and privacy-respecting content management systems.

The synthetic data generator generated 10,000 JSON-formatted records representing diverse access scenarios adhering to these policies. Each record contains:\\
\textbf{Request Context:} User identifier, requested action, target resource, timestamp \\
\textbf{System State:} Resource workflow state (submitted, graded), review count, user relationships (authorship, reviewer history) \\
\textbf{Decision Label:} Binary allow/deny decision computed from policy evaluation \\
\textbf{Explanation:} Natural language rationale tracing the decision to specific policy conditions. \\
Due to computational resource constraints in the experimental environment, we utilized 10\% of the generated dataset for model training while maintaining balanced representation across all seven policy types and action categories. 

\section{Results and Discussion}
\label{sec:results}
\begin{figure}[!ht]
    \includegraphics[width=1\columnwidth]{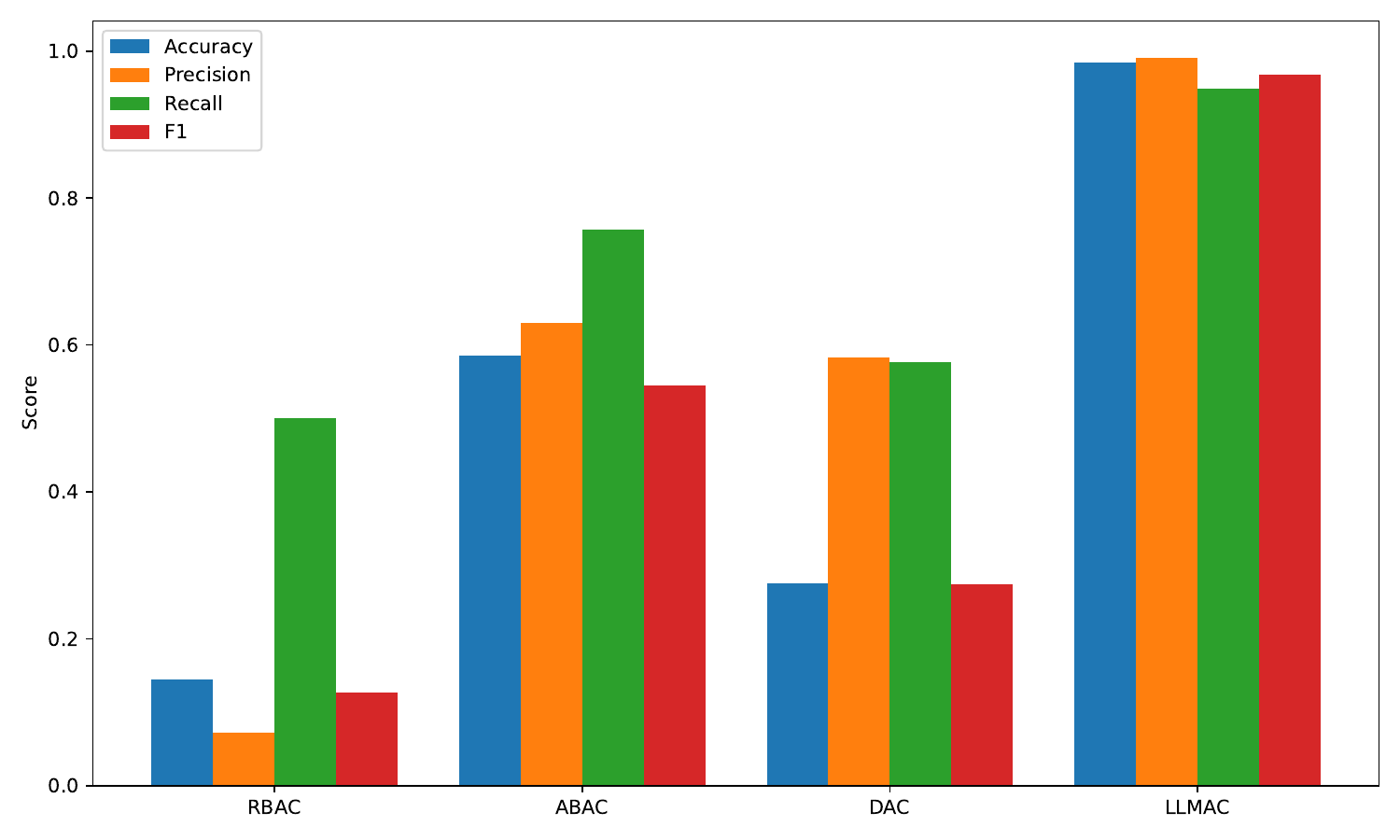}
    \centering
    \caption{Performance Comparison of Access Control Models}
    \label{fig_performance_comparison}
    \vspace{-10pt}
\end{figure}

Our proposed LLM-based access control model (LLMAC) shows outstanding results across all evaluation metrics, achieving 98.5\% accuracy, 99.1\% precision, 94.8\% recall, and 96.8\% macro-F1. These results indicate a balanced ability to approve legitimate requests and block unauthorized ones. This is essential for securing sensitive information and maintaining privacy in dynamic workflows. We have compared the test dataset with the traditional access control model as shown in Figure \ref{fig_performance_comparison}.

\vspace{-10pt}

According to performance comparison in Figure \ref{fig_performance_comparison}, traditional models perform noticeably worse under the same policy load. ABAC reaches 58.5\% accuracy (macro-F1 54.5), DAC attains 27.5\% accuracy (macro-F1 27.5), and RBAC performs poorly except in very simple cases. This highlights the limitation of static rule-based methods in handling context-dependent access decisions. Table \ref{tab:per_action_accuracy} shows action-wise accuracy comparison between these models.

\vspace{-5pt}
\begin{table}[htbp]
    \centering
    \caption{Action-wise Accuracy Comparison}
    \label{tab:per_action_accuracy}
    \begin{tabular}{|p{2.8cm}|p{1cm}|p{1cm}|p{1cm}|p{1cm}|}
        \hline
        \textbf{Action} & \textbf{RBAC} & \textbf{ABAC} & \textbf{DAC} & \textbf{LLMAC} \\ 
        \hline
        upload\_homework       & 1.000 & 1.000 & 1.000 & \textbf{1.000} \\
        \hline
        replace\_homework      & 0.079 & 0.079 & 0.079 & \textbf{0.921} \\
        \hline
        submit\_homework       & 0.000 & 1.000 & 0.000 & \textbf{1.000} \\
        \hline
        review\_homework       & 0.000 & 1.000 & 0.000 & \textbf{1.000} \\
        \hline
        revise\_review         & 0.000 & 1.000 & 1.000 & \textbf{1.000} \\
        \hline
        grade\_homework        & 0.000 & 0.000 & 0.000 & \textbf{1.000} \\
        \hline
        append\_review\_to\_grade & 0.000 & 0.000 & 0.000 & \textbf{1.000} \\
        \hline
    \end{tabular}
    \vspace{-10pt}
\end{table}

The comparison in Table \ref{tab:per_action_accuracy} shows a noticeable pattern for important security tasks where protecting content and keeping workflows secure is very important. For actions that anyone can do, such as uploading homework, all methods get 100\% accuracy, which is expected. However, the actions that need more complex reasoning, for instance, replacing homework (which checks authorship and whether it is unsubmitted or not), grading homework (which requires a minimum review), and appending a review to a grade (which links different users), the LLMAC stays very accurate (92–100\%). Rule-based systems often fail or drop to zero since they can not handle time-based rules, dynamic duty separation, or relationship checks. ABAC works well for some simpler actions, such as submitting or reviewing homework, while its accuracy drops when actions need multiple workflow conditions and quota limits to be checked together.

These results demonstrate that LLMAC can combine the strengths of role-based, attribute-based, and ownership-based access controls into a single model. It also provides clear explanations for its decisions, which helps with audits and following rules. The natural language explanations make the system easier to understand, which is important for privacy, investigating problems, and responsible management. Being able to see why an action is allowed or denied also lowers risks, helping administrators stop threats such as impersonation, collusion, and unauthorized changes to content. Figure \ref{fix_explanation_output} represents some test cases illustrating how the model explains its decisions in realistic scenarios:

\vspace{-8pt}
\begin{figure}[!ht]
    \includegraphics[width=1\columnwidth]{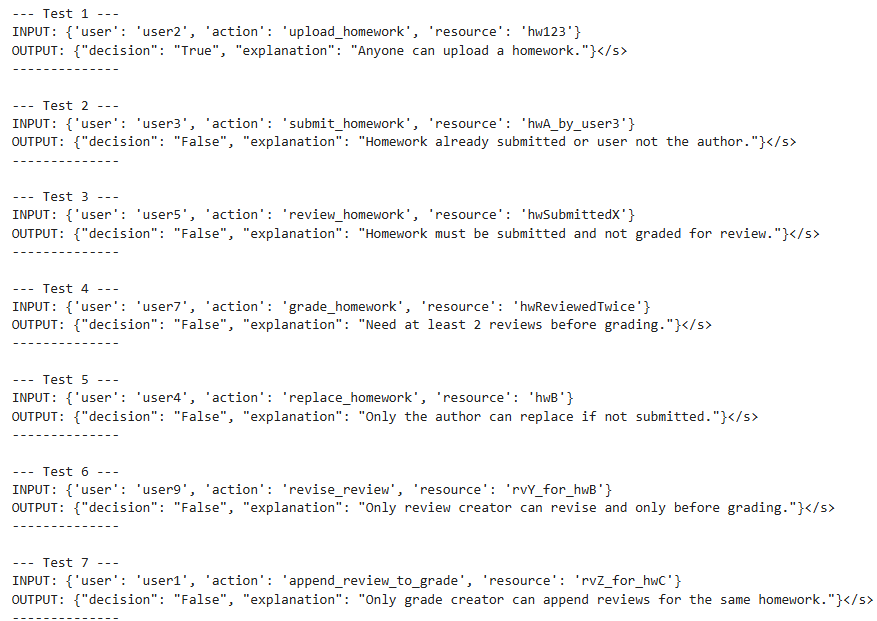}
    \centering
    \caption{LLMAC Decision with Explanation}
    \label{fix_explanation_output}
    \vspace{-20pt}
\end{figure}




\section{Conclusion}
\label{sec:conclusion}

In this research, we presented an innovative access control framework, LLMAC, based on large language model (LLM) that unifies traditional models such as RBAC, ABAC, and DAC into a single, explainable system. Our approach demonstrates significant improvements in authorization accuracy and robustness. Our LLM-based model helps protect sensitive content and privacy by providing consistent decisions along with explanations that are easy for humans to understand. These explanations also help security operations by making incident investigations easier, reducing insider threats~\cite{zisad2024towards}, and building user trust, all key aspects of reliable cybersecurity solutions. The results show that the explainable approach we designed can effectively address the nuanced, dynamic, and context-dependent policies that are required in modern workflows, which traditional models struggle to enforce reliably. However, the LLM-based model always has higher inference latency compared to lightweight rule engines. This trade-off is acceptable since it provides much better accuracy and clearer, understandable decisions. In practice, techniques such as quantization, batching, and scheduling can reduce response time while maintaining the model’s accuracy and explainability~\cite{10.5555/3737916.3739798}. This approach also aligns with many enterprise scenarios where security and privacy take precedence over microsecond latency. The future work will be assessing the security threats to the model~\cite{zisad2025threatgpt}, such as prompt injection and data leakage, and integrating risk‑adaptive, context‑aware signals with formal verification to strengthen policy conformance across domains.

\balance
\bibliographystyle{IEEEtran}
\bibliography{references}

\end{document}